\documentclass[twocolumns]{aa} 
\usepackage{graphicx,natbib}
\usepackage{graphicx,color}
\usepackage{inputenc}
\usepackage{txfonts}
\usepackage{longtable}

\newcommand{\Msun}{M$_\odot$ }
\newcommand{\Yini}{Y$_\mathrm{ini}$ }
\newcommand{\DY}{$\Delta$Y$_\mathrm{ini}$ }
\newcommand{\ig}[1]{} 

\begin{document} 

   \authorrunning{Charbonnel \& Chantereau}
   \titlerunning{}

   \title{	
	Evolution of long-lived globular cluster stars}
   \subtitle{II. Sodium abundance variations on the asymptotic giant branch as a function of globular cluster age and metallicity}

   \author{Corinne Charbonnel\inst{1,2} \& William Chantereau\inst{1}}

   \institute{Department of Astronomy, University of Geneva, Chemin des Maillettes 51, 1290 Versoix, Switzerland\\
   \email{corinne.charbonnel@unige.ch} \and
   Institut de Recherche en Astrophysique et Plan\'etologie, CNRS UMR 5277, Universit\'e de Toulouse, 14, Av.E.Belin, 31400 Toulouse, France\\
             }

  \date{Accepted for publication}

  \abstract
  {Long-lived stars in globular clusters exhibit chemical peculiarities with respect to their halo counterparts. In particular, sodium-enriched stars are identified as belonging to a second stellar population born from cluster material contaminated by the hydrogen-burning ashes of a first stellar population. Their presence and numbers in different locations of the colour-magnitude diagram provide important constraints on the self-enrichment scenarios.  In particular, the ratio of Na-poor to Na-rich stars on the asymptotic giant branch (AGB) has recently been found to vary strongly from cluster to cluster (NGC~6752, 47~Tuc, and NGC~2808), while it is relatively constant on the red giant branch (RGB).}
  {We investigate the impact of both age and metallicity on the theoretical sodium spread along the AGB within the framework of the fast rotating massive star (FRMS) scenario for globular cluster self-enrichment. }
  {We computed evolution models of low-mass stars for four different metallicities ([Fe/H]=-2.2, -1.75, -1.15, -0.5) assuming the initial helium-sodium abundance correlation for second population stars derived from the FRMS models and using mass loss prescriptions on the RGB with two realistic values of the free parameter in the Reimers formula.}
  {Based on this grid of models we derive the theoretical critical initial mass for a star born with a given helium, sodium, and metal content that determines whether that star will climb or not the AGB. This allows us to predict the maximum sodium content expected on the AGB for globular clusters as a function of both their metallicity and age. 
We find that 
(1) at a given metallicity, younger clusters are expected to host AGB stars exhibiting a larger sodium spread than older clusters and
(2) at a given age, higher sodium dispersion along the AGB is predicted in the most metal-poor globular clusters than in the metal-rich ones. We also confirm the strong impact of the mass loss rate in the earlier evolution phases on the Na cut on the AGB: the higher the mass loss, the stronger the trends with age and metallicity. }
  {The theoretical trends we obtain provide, in principle, an elegant qualitative explanation to the different sodium spreads that are observed along the AGB in the Galactic globular clusters of different ages and [Fe/H] values. Although it is real, the slope with both age and metallicity is relatively flat, although it steepens when accounting for mass loss variations. Therefore, additional parameters may play a role in inducing cluster to cluster variations, that are difficult to disentangle from existing data.}
    
\maketitle

%

\section{Introduction}

The discovery of large sodium (Na) abundance variations from star to star both at the main sequence (MS) turnoff and on the red giant branch (RGB) within Galactic globular clusters (GC; e.g. \citealt{CottrellDaCosta1981,Gratton01}) has revolutionized the classical paradigm that describes these systems as large aggregates of coeval stars formed out of homogeneous material.
The ubiquitous presence of the Na-O anticorrelation in Galactic and extragalactic GCs \citep[e.g.][]{carretta09a,Mucciarellietal2009,Lind11,Larsenetal14}, together with the C-N and the Mg-Al anticorrelations \citep[see the review by][ and references therein]{Gratton12}, are interpreted as a result of self-enrichment of these systems during their early evolution.
On the one hand, GC stars with chemical properties similar to those of Galactic field stars are thought to have formed from original proto-cluster material, and are often referred to as first population (or first generation, hereafter 1P) GC stars. On the other hand, GC stars that exhibit different degrees of Na, N, and Al enrichment anticorrelated with O, C, and Mg depletion (so-called second population, 2P, or second generation stars) must have formed from original gas polluted to various degrees by hydrogen-burning ashes  
ejected by more massive, short-lived, 1P GC stars \citep[e.g.][]{Prantzos07}. 

The frontier between these different chemical populations and their generational connections within individual GCs are far from being clear. Several GC self-enrichment scenarios are competing to explain the observations. Each of them invokes specific ``polluters"\footnote{Fast rotating massive stars (FRMS, masses above 25~M$_{\odot}$; \citealt{Prantzos06,Decressin07a,Decressin07b,Krause13}), asymptotic giant branch stars (AGB, masses of $\sim 6.5$~M$_{\odot}$; \citealt{Ventura01,Ventura13,D'ercole10}), massive binary stars (masses between $\sim$ 10 and 20~M$_{\odot}$; \citealt{DeMink09}), or supermassive stars (masses $\sim 10^4$~M$_{\odot}$; \citealt{DenissenkovHartwick14}).} that imply very different conditions and timelines for the formation of GC stars with the observed chemical peculiarities. While the formation of the two populations is separated by $\sim$ 50 - 100~Myr in the case of the AGB polluter scenario \citep[e.g.][]{D'ercole10}, a limited delay lower than $\sim$ 3.5 to 8~Myr is expected in the FRMS polluter scenario \citep[e.g.][]{Krause13}. 
Additionally, different assumptions can be made for the stellar initial mass function (IMF) of the different populations. 
In most cases, ``classical" IMF slopes (e.g., Salpeter or Kroupa IMF) are considered for the 1P polluters. This has extreme consequences on the determination of the clusters' IMF, as this implies the loss of most of 1P stars ($\sim 95 \%$ ) towards the Galactic halo in order to explain the observed number ratio of low-mass stars along the O-Na distribution \citep[the so-called mass budget problem, see e.g][]{Norris04,Prantzos06,Carretta10,SchaererCharbonnel2011}.
However, \citet{Charbonnel14} propose one variation of the FRMS scenario where only massive stars form from pure original proto-cluster material, and all low-mass GC stars form from a mixture between H-processed stellar ejecta and proto-cluster material 
in variable proportions accounting for both the observed Li-Na anti-correlation \citep[e.g.][]{Lind09} and the typical percentage of low-sodium stars that are usually classified as 1P stars ($\sim 30 \%$; \citealt{Prantzos06,carretta09b,Carretta13}). This strongly alleviates the mass budget issue, in better agreement with constraints provided by young massive star clusters (\citealt{Bastianetal14,Hollyhead15}; Krause et al., submitted) and star counts in the halo of dwarf galaxies \citep{Larsenetal14}.

Despite their differences,  all the self-enrichment scenarios predict that the 2P GC stars that are enriched in sodium were born with higher helium content than 1P stars, as helium is the main product of hydrogen burning. 
This has been confirmed by direct spectroscopic measurements for a few horizontal branch (HB) stars in the GC NGC 2808 \citep{Marino14}. 
The implication of initial helium enhancement on the HB morphology has been extensively discussed in the literature \citep[e.g.][ and references therein]{D'Antona10,Chantereau15}. 
Helium variations are also invoked to explain (at least partly) the peculiar photometric patterns (multiple sequences and/or spreads at the turnoff and/or along the subgiant and giant branches) in the colour-magnitude diagram (CMD) of several GCs \citep{Milone12,Marino12,Piotto12,Monelli13}.
However, the degree of He enrichment with respect to that of Na strongly varies from one potential polluter to the 
other\footnote{In the AGB scenario, all 2P stars spanning a large range in Na are expected to have very similar He (maximum $\sim$ 0.36 - 0.38 in mass fraction). This is due to the fact that He enrichment of the envelope of intermediate-mass stars results from the second dredge-up on the early-AGB, which is independent of the mass of the AGB progenitor \citep[e.g.][]{Forestini97,Doherty14} and occurs before the TP-AGB where hot bottom burning might affect the abundances of Na. 
On the other hand in the FRMS scenario the Na and He enrichment are correlated as they both result from hydrogen-burning on the main sequence (see \citealt[][ hereafter Paper I]{Chantereau15}, and references therein).}.

In this series of papers we investigate the impact of helium enrichment associated with the other abundance peculiarities in Na, O, Mg, Al, C, and N on the properties of low-mass GC stars all along their evolution, and the implications on GC CMD. 
In Paper I we presented a large grid of stellar evolution models for low-mass stars (initial masses between 0.3 and 1.0~M$_{\odot}$) with an initial helium mass fraction between 0.248 and 0.8 computed for [Fe/H]=-1.75 (close to the value of one of the best studied GC, NGC 6752). 
With these models we explored the implications of an extreme helium spread on the main stellar properties  (evolution paths, lifetimes, chemical nature of the white dwarf remnants, and mass and helium content of stars in different areas of the Hertzsprung-Russel diagram, HRD). 
In \citet[][ hereafter C13]{Charbonnel13}, we used this grid of stellar models to interpret the apparent lack of Na-rich AGB stars in NGC 6752, which was considered a serious problem for stellar evolution theory when it was discovered \citep{Campbell13}. We showed in particular that the He-Na correlation needed to explain the AGB data in this GC corresponds to the one predicted by the FRMS scenario, but that it is at odds with the le lack of correlation predicted in the AGB pollution scenario. 

Here we complete the study of C13. We investigate in particular the impact of both metallicity and age on the spread in Na expected along the AGB for He-enriched stars, when assuming the initial He-Na correlation resulting from the FRMS model and Reimers mass loss prescription (with two different values of the $\eta$ parameter) for low-mass evolved stars. 
In \S~\ref{section:models} we present the input physics of the low-mass stellar models we have computed for the present purpose, which cover the range in [Fe/H] between -2.2 and -0.5. 
In \S~\ref{section:0.75} we describe how helium and metallicity affect the passage on the AGB for a star of given mass (we focus on the example of a 0.75~M$_{\odot}$ model). 

We use our grid of models in \S~\ref{section:Napredictions} to derive the critical initial mass for a star born with a given initial helium mass fraction and [Fe/H] to climb or not the AGB. This allows us to draw a theoretical AGB / no-AGB limit as a function of initial stellar mass and helium content for four values of [Fe/H] (-2,2,-1.75, -1.15, and -0.5).  We then derive the theoretical sodium spread expected along the AGB as a function of GC metallicity and age (between 9 and 13.4~Gyr, which covers the age range of Galactic GCs). We discuss open issues, and compare the 
theoretical predictions to observations in \S~\ref{section:conclusion}. 

\ig{One observes different delta [Na/Fe] on the AGB of globular clusters, we want to investigate this phenomenon in the framework of the FRMS scenario and extend the previous work we have done before in the letter [] to different metallicities. FRMS a presenter ds l'intro}

\section{Stellar evolution models. Assumptions}
\label{section:models}

\subsection{Basic input physics}
\label{subsec:inputmicrophysics}
We compute evolution models of low-mass stars with the code STAREVOL assuming the same input physics as in Paper I and C13. 
We refer to these papers for references and details on the equation of state, opacities, nuclear reaction rates, and mixing length parameter for convection. 

Mass loss is accounted for by following the \citet{Reimers75} prescription up to the end of central He-burning with the $\eta$ parameter equal to 0.5 (i.e., close to the median value of 0.477 $\pm$ 0.070 for GC RGB stars according to \citealt{McDonald15}), 
and following the \citet{Vassiliadis93} prescription in the more advanced phases. 
For the two  extreme metallicities of the present grid ([Fe/H] = -2.2 and -0.5) we investigate the impact of a higher $\eta$ parameter (0.65, maximum value inferred by \citealt{McDonald15}) in the Reimers formula for the stellar models that are close to the 
AGB / no-AGB limit (\S~\ref{sec:AGB-noAGBlimit}).

We do not include atomic diffusion, rotation, or overshooting in the computations. 
Models are built for four values of [Fe/H] (-2.2, -1.75, -1.15, and -0.5).
The chemical composition we assume for the 1P and 2P models is described below. 
The mass range and the step in mass of the models of the grid are given in \S~\ref{subsec:curveAGBlimit}.

\subsection{Chemical composition of first population stars}
\label{subsec:firstpopmodels}

The initial helium mass fraction \Yini of 1P GC stars at a given metallicity is obtained with 
the relation \Yini = Y$_0$ + ($\Delta$Y / $\Delta$Z) $\times$ Z, where Z is the heavy element mass fraction.
The primordial helium mass fraction Y$_0$ chosen is equal to 0.2479 \citep{Coc04}. 
The slope $\Delta$Y / $\Delta$Z is derived from the solar\footnote{For the solar value we use the helium mass fraction obtained by calibrating a 1~M$_{\odot}$, Z$_{\odot}$ standard model in both luminosity and radius at the age of 4.57~Gyr, using the \citet{Grevesse93} mixture as in our present computations.} and primordial helium abundances and is equal to 1.62. 
We include $\alpha$-enhancement depending on [Fe/H] following \cite{Carretta10}. 
For the other elements we adopt a solar-scaled composition using \citet{Grevesse93} to be consistent with the chemical composition of the FRMS models of \citet{Decressin07a}. 
The corresponding values for Z, [$\alpha$/Fe], and \Yini are given in Table~\ref{table1}. 

\begin{table}[ht]
	\centering 
	\begin{tabular}{c  c  c  c }       
	\hline 
	[Fe/H] & Z & [$\alpha$/Fe] & \Yini (1P)\\ \hline
	-0.5 &  $7.9 \times 10^{-3}$ & +0.2 & 0.255 \\
	-1.15 & $2.1 \times 10^{-3}$ & +0.3 & 0.249 \\
	-1.75 & $5.4 \times 10^{-4}$ & +0.3 & 0.248 \\
	-2.2 &  $2.1 \times 10^{-4}$ & +0.35 & 0.248 \\
	\hline 
	\smallskip
	\end{tabular} 
\caption{Heavy element mass fraction Z, $\alpha$-enrichment, and helium mass fraction of the first population stars for the different values of [Fe/H] of the present grid of stellar models} 
\label{table1}
\end{table}

\subsection{Chemical composition of second population stars}
\label{subsec:secondpopmodels}

For the initial chemical composition of 2P stars we adopt the correlation between He and Na abundances predicted by the FRMS models of \citet{Decressin07b} accounting for dilution; the abundances of C, N, O, Mg, and Al are scaled accordingly. We refer to Paper I and C13 for details (see e.g. Fig.1 in C13 and in Paper I). 
For each value of [Fe/H], we compute models of 2P stars with \Yini equal to 0.26, 0.27, 0.3, 0.33, 0.37, 0.4, 0.425, and 0.45. 
This last value corresponds to a Na enrichment $\Delta$[Na/Fe] of 0.84~dex with respect to  1P stars. In Paper I, 2P models were computed with \Yini as high as 0.8.
However, when the initial He mass fraction is higher than 0.5, stars that reach the AGB phase have lifetimes that are shorter than the age of the youngest Galactic GCs. Therefore such models are not relevant for the purpose of the present study and are not presented here, except for the specific and illustrative case of the 0.75~M$_{\odot}$ models discussed in \S~\ref{section:0.75}.

\subsection{A definition of the AGB / no-AGB limit} 
\label{sec:AGB-noAGBlimit}

In this paper we consider two different criteria for a star to be counted as an AGB.
We first take the same AGB criterion as in C13, i.e. we count both thermally pulsing AGB (TP-AGB) stars and those stars that make an excursion on the early-AGB without necessarily undergoing thermal pulses (TP) before moving towards the planetary nebulae phase. As a second criterion, we  consider that a star can be counted as an AGB only when the corresponding model undergoes at least two helium thermal pulses (TP) after the end of the early-AGB, which is a more conservative definition.
Since the lifetime of the TP-AGB is much shorter than the total duration of the AGB sequence \citep[i.e., between the end of central He-burning and the end of the last TP; e.g.][]{Forestini97,Herwig05,Karakas05,Marigoetal13}, we are aware that this second criterion might be too strict from the point of view of the star counts for low-luminosity AGB stars. However, we introduce it in order to avoid contamination by RGB stars in star counts on the upper AGB. 

\section{Impact of initial helium and metal content on a 0.75~M$_{\odot}$ star}
\label{section:0.75}

The influence of the initial composition of a star on its evolution and fate is described at length in the literature.
In Paper I, we quantified for the first time how very high initial helium content (\Yini between 0.4 and 0.8) modifies both the stellar lifetime and the evolution path within the HRD for a specific value of  [Fe/H] equal to -1.75.
Here we note the main points and the dependency with [Fe/H]. 
For illustration purposes, we present theoretical predictions for the 0.75~M$_{\odot}$ case for which we computed the four considered metallicity models with three different initial helium mass fractions: \Yini(1P), and \Yini(1P)+$\Delta$\Yini, $\Delta$\Yini being chosen equal to 0.152 and 0.552 for the sake of homogeneity.  
The definitions of the different evolution phases are the same as in Paper I (see their \S~4), except for the additional and more conservative limit we adopt here (\S~\ref{sec:AGB-noAGBlimit}) for the AGB phase. 

\begin{figure}[ht]
\begin{center}
\includegraphics[width=.47\textwidth]{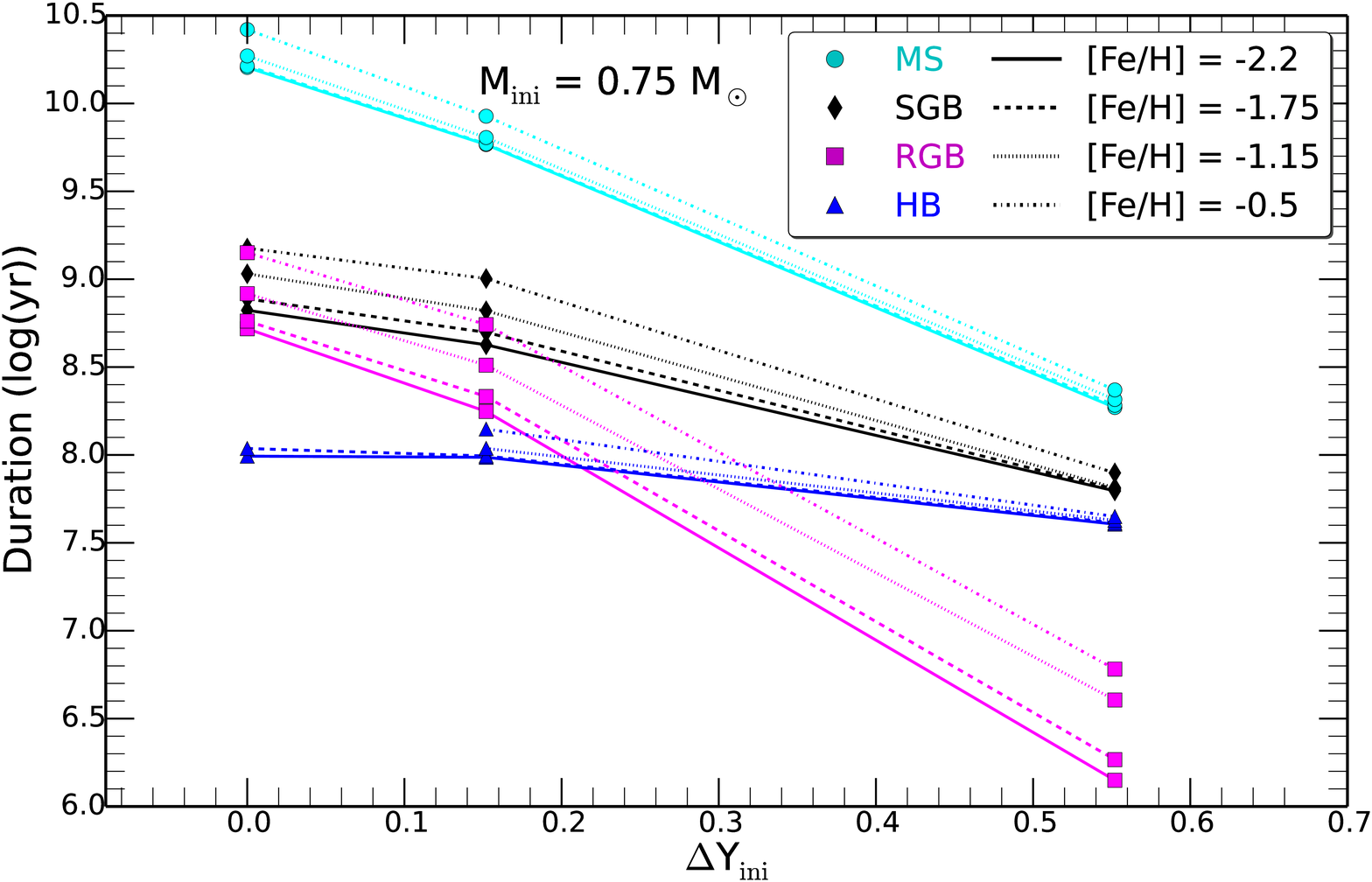} 
\end{center}
\caption{Duration of the different evolution phases up to the end of central helium burning for the 0.75 \Msun models as a function of initial helium enrichment for the four different values of [Fe/H] (see colours and symbols in inset; MS = main sequence; SGB = subgiant branch; RGB = red giant branch; HB = horizontal branch). The abscissa $\Delta$\Yini is the difference between the initial helium mass fraction of a given model and that of 1P stars at the considered metallicity} 
\label{Fig:Lifetime}
\end{figure}

\subsection{Stellar lifetime}

Helium enrichment in the initial chemical composition of 2P stars reduces the Thomson scattering opacity and increases the mean molecular weight within the stellar interior with respect to the 1P case. This modifies the evolution path in the HRD; helium-rich stars of a given initial mass evolve at a higher effective temperature and luminosity than their helium-normal counterparts. These effects are reinforced when [Fe/H] decreases, as this further decreases stellar opacity.
Consequently, the duration of the MS, the subgiant branch (SGB), and the RGB are shortened when higher \Yini and lower [Fe/H] are considered. This is quantified for the 0.75M$_{\odot}$ models in Fig.~\ref{Fig:Lifetime} where we plot the duration of the different evolution phases as a function of initial He enrichment  for the considered [Fe/H] values. 
For the age range covered by Galactic GCs (typically between $\sim$ 9 and 13~Gyr), the only 0.75M$_{\odot}$ stars that are still active are those with relatively low helium enrichment, while all the others have already finished their lives and remain only in the form of white dwarfs (see below).

\subsection{Central helium burning}

\subsubsection{To ignite (or not) central helium burning and the nature of the white dwarf remnants}
\label{subsubsub:WD}

The domain in initial stellar mass and helium mass fraction 
where stars ignite central helium burning is discussed in Paper I for [Fe/H] equal to -1.75 (see their Fig.~16): The higher \Yini, the lower the critical initial stellar mass for a star to burn helium. Additionally for a given \Yini, the minimum stellar mass to ignite helium decreases with metallicity. Therefore, changing [Fe/H] induces a shift in the (M; \Yini) limit for a star to finish its life as an He or CO white dwarf (WD).  

For the 0.75~M$_{\odot}$ case, the 1P models (i.e. $\Delta$\Yini=0) computed with the highest [Fe/H] values (-0.5 and -1.15)
actually end their lives as helium WDs, whereas those computed with the lowest metallicities ignite helium in their core at the tip of the RGB ([Fe/H] =-2.2), or while crossing the HRD from the RGB tip towards high effective temperature (hot-flasher, [Fe/H] = -1.75). 
For the two other values of $\Delta$\Yini assumed here, all the 0.75~M$_{\odot}$ models reach the conditions to burn helium whatever the [Fe/H] value and therefore finish their lives as CO WD  (Fig.~\ref{masseWD}).

\begin{figure}
\begin{center}
\includegraphics[width=.47\textwidth]{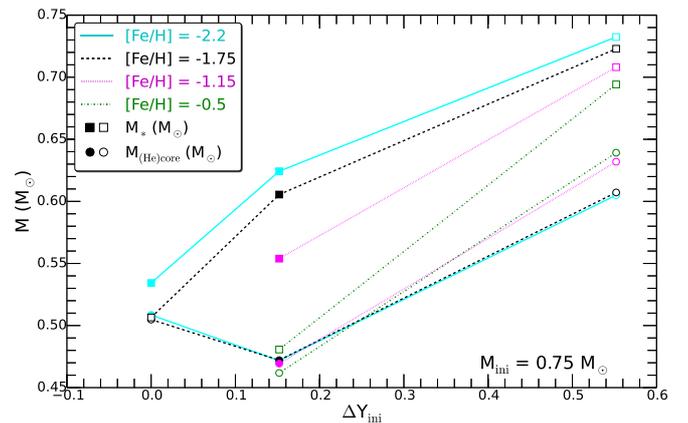} 
\end{center}
\caption{Total stellar mass and mass of the helium core (squares and circles respectively) of the 0.75 \Msun models at the arrival on the ZAHB as a function of \DY and for the different metallicities (colour-coded, see inset). For the highest [Fe/H] values (-1.15 and -0.5), the 1P models (i.e. $\Delta$\Yini=0) do not ignite central He burning and therefore do not appear on the plot. Full and open symbols correspond respectively to the models that do or do not reach the TP-AGB phase (and undergo at least two thermal pulses) after central helium burning}
\label{ZAHB}
\end{figure}

For a given initial stellar mass, central helium ignition occurs at lower luminosity on the RGB when \Yini increases, as a result of lower degeneracy and higher temperature of the helium core. Additionally, the total stellar mass at the time of central helium ignition is higher when higher \Yini values are considered, as shown in Fig.~\ref{ZAHB} for the 0.75 \Msun models. 
This results from more modest amount of mass lost by the stellar wind on the RGB, due to shorter RGB lifetime and lower luminosity reached at the RGB tip when \Yini increases. Again, all these effects are reinforced when [Fe/H] is lower.

\subsubsection{Modes of helium ignition and properties on the horizontal branch}

The mode of helium ignition for a given initial stellar mass depends on the 
combination between \Yini and [Fe/H] that both affect the degree of degeneracy of the helium core. 
On the one hand, for the highest helium enrichment considered here ($\Delta$\Yini= 0.552) all the 0.75~M$_{\odot}$ models ignite helium in non-degenerate conditions independently of the metallicity we assume. In this regime, the stars arrive on the zero age horizontal branch (ZAHB) with both higher total mass and core mass (and as a consequence with higher effective temperature) when \Yini increases. 
On the other hand, when the initial helium content is lower ($\Delta$\Yini= 0 and 0.152) the 0.75~M$_{\odot}$ models that ignite helium undergo the helium flash at the RGB tip (or while crossing the HRD); they arrive on the ZAHB with a lower effective temperature when \Yini increases, due to their higher total mass and lower ratio between the core and the envelop masses. The dependency with metallicity of these effects on the mass of the star and of its core on the ZAHB is illustrated in Fig.~\ref{ZAHB}.

Finally, the duration of the central helium burning phase decreases with metallicity, and it varies by $\sim$ 63 \% over the investigated range in $\Delta$\Yini for the 0.75~M$_{\odot}$ case (Fig.~\ref{Fig:Lifetime}). 

\subsection{To become or not an AGB}
\begin{figure}
\begin{center}
\includegraphics[width=.47\textwidth]{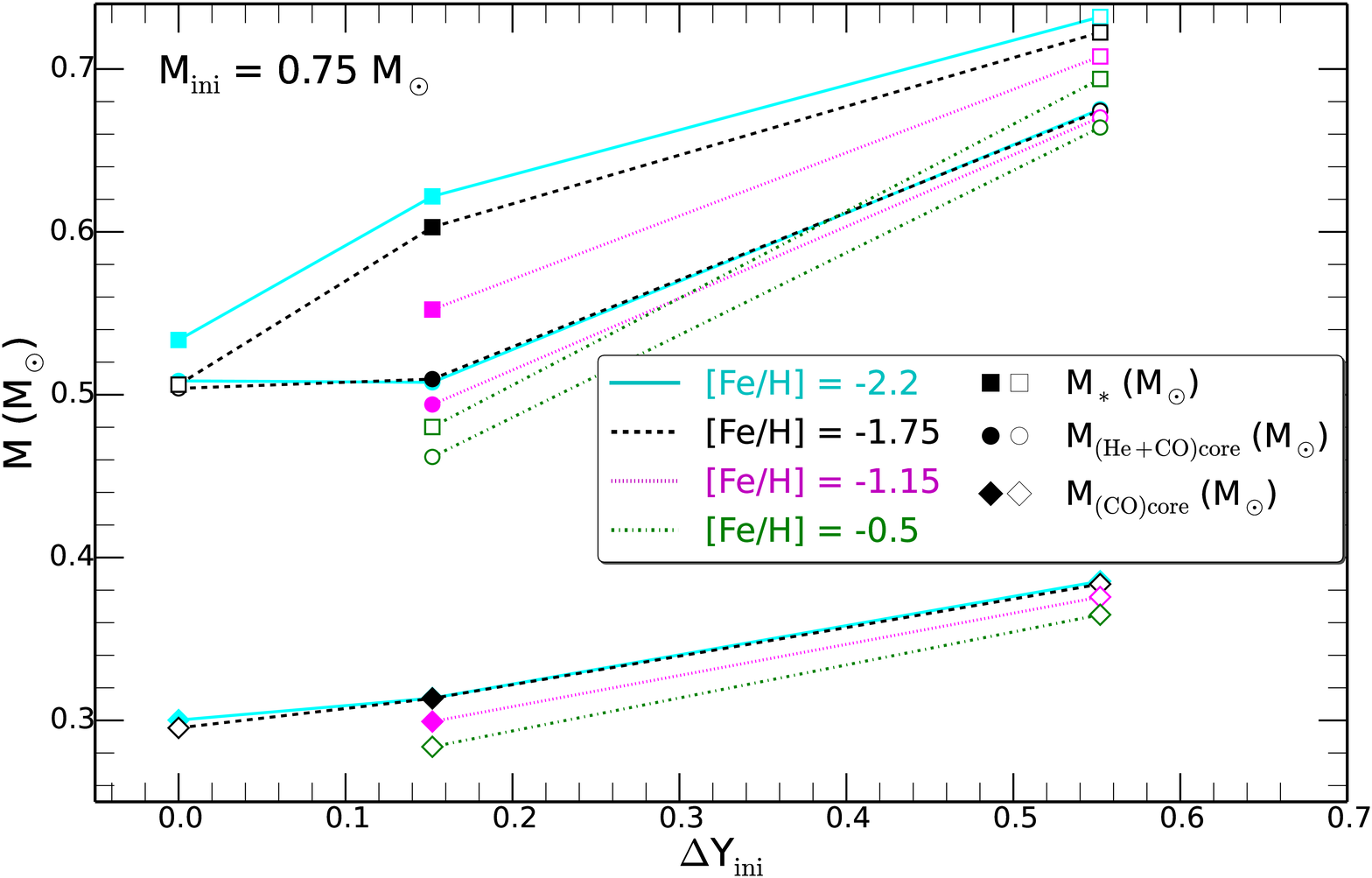} 
\end{center}
\caption{Total stellar mass, mass of the (He+CO) core, and mass of the CO core (squares, circles, and diamonds respectively) at the end of HB for the 0.75 \Msun models at the different metallicities (colour-coded) as a function of \DY. Full and open symbols correspond respectively to the models that do or do not reach the TP-AGB phase (and undergo at least two thermal pulses) after central helium burning}
\label{EndHeb}
\end{figure}

At the end of central helium burning, a minimum envelope mass is required for a star to climb the AGB and to undergo thermal pulses (TP). Additionally, the number of TPs depends on the mass of the stellar layers located above the helium-burning shell (HeBS). 
For the 0.75~M$_{\odot}$ models the relevant information can be retrieved from Fig.~\ref{EndHeb} where we plot (at the end of central helium burning) the mass of the CO core, the mass of the CO+HeBS region, and the total remaining mass of the star as a function of helium enrichment for the different values of [Fe/H]. In addition, we use different symbols depending on whether or not a given model becomes an AGB (filled and open symbols, respectively) according to the more conservative definition (i.e., at least two thermal pulses). 
All of these quantities increase with increasing helium enrichment and decreasing metallicity. Consequently the mass of the layers above the HeBS decreases for very high \Yini (above 0.4) and lower [Fe/H].

As a consequence of this behaviour, all the moderately helium-enriched 0.75~M$_{\odot}$ models ($\Delta$\Yini=0.152) climb the TP-AGB except for the most metal-rich one ([Fe/H] = -0.5) that has an envelope above the HeBS that is too small (in mass) because its longer RGB lifetime enables more mass loss from the wind. 
However, all the most helium-rich  0.75~M$_{\odot}$ models ($\Delta$\Yini=0.552) become AGB-manqu\'e (or undergo only one TP) independently of [Fe/H]. In this case all the models develop much more massive CO cores than in the helium-normal case; therefore, they leave the HB with a very narrow stellar envelope that cannot be swollen by the energy released by the contraction of the core and nuclear shell burning after central He-burning.  This actually prevents the stars from climbing the TP-AGB.

\subsection{Remnants}
\begin{figure}
\begin{center}
\includegraphics[width=.47\textwidth]{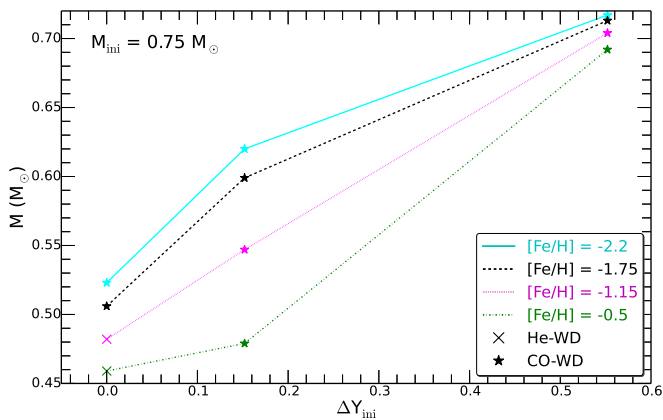} 
\end{center}
\caption{Mass of the white dwarf remnant for the 0.75 \Msun models as a function of \DY for the different metallicities (colour-coded). Crosses and stars correspond respectively to helium and CO white dwarfs}
\label{masseWD}
\end{figure}

The nature (He or CO) and the mass of the white dwarfs at the end of the evolution of the different 0.75 \Msun models are presented in Fig.~\ref{masseWD}. As already discussed in \S~\ref{subsubsub:WD}, He WD are obtained only for the canonical helium abundance and for the most metal-rich initial composition. Even so, these two models have stellar lifetimes that are longer than the Hubble time (Fig.~\ref{Fig:Lifetime}). Therefore no He WDs with 0.75 \Msun progenitors are expected to be present today in GCs. However and as discussed in Paper I (see their Fig.16), He WDs are expected to form on much shorter lifetimes for slightly higher initial mass and helium content. 
Finaly, we note that the theoretical remnant mass strongly increases with both increasing initial helium content and decreasing metallicity, as a result of modified stellar evolution as described above (see  Fig.~\ref{masseWD}). 

\section{Sodium spread on the AGB - Theoretical trends with age and metallicity for the current model assumptions}
\label{section:Napredictions}

As discussed in \S~\ref{section:0.75} (see also Paper I and C13), helium-enriched stars have shorter lifetimes for a given initial stellar mass than helium-normal stars; this adds to the impact of initial metallicity on the duration of the different evolution phases. 
Additionally, above a cut-off value of the initial helium abundance that also depends on metallicity, stars do not climb the AGB and evolve directly to the WD sequence at the end of central helium burning.

Importantly in the framework of the FRMS scenario for GC self-enrichment a helium-sodium correlation is predicted in the initial composition of 2P stars (\S~\ref{subsec:secondpopmodels}). As a consequence, a cut-off value is also expected for the sodium abundance of GC AGB stars, as already discussed in C13 for the specific case of NGC~6752, and as quantified below as a function of age and metal content.

\subsection{Critical (Y$_{\rm ini}$; M$_{\rm ini}$; [Fe/H]) combination for a star to become an AGB}
\label{subsec:curveAGBlimit}

In order to pinpoint the critical initial mass at which a model fails to climb the AGB for each (Y$_{\rm ini}$; [Fe/H]) combination, we first computed a grid of 2P models with the assumptions presented in \S~\ref{section:models}; the initial masses varied between 0.6 and 0.95~M$_{\odot}$ with a mass step of 0.05~M$_{\odot}$.  We then computed additional models with a refined mass step of 0.01~M$_{\odot}$ around the AGB / no-AGB limit. 
The critical mass at which a star becomes an AGB is then obtained by interpolating between the closest models that become AGB or not (according to both AGB criteria).  In addition,  for the two extreme [Fe/H] values we also derive the AGB / no-AGB limit using different values for the $\eta$ parameter in the Reimers prescription on the RGB.

\begin{figure*}
\begin{center}
\includegraphics[width=0.87\textwidth]{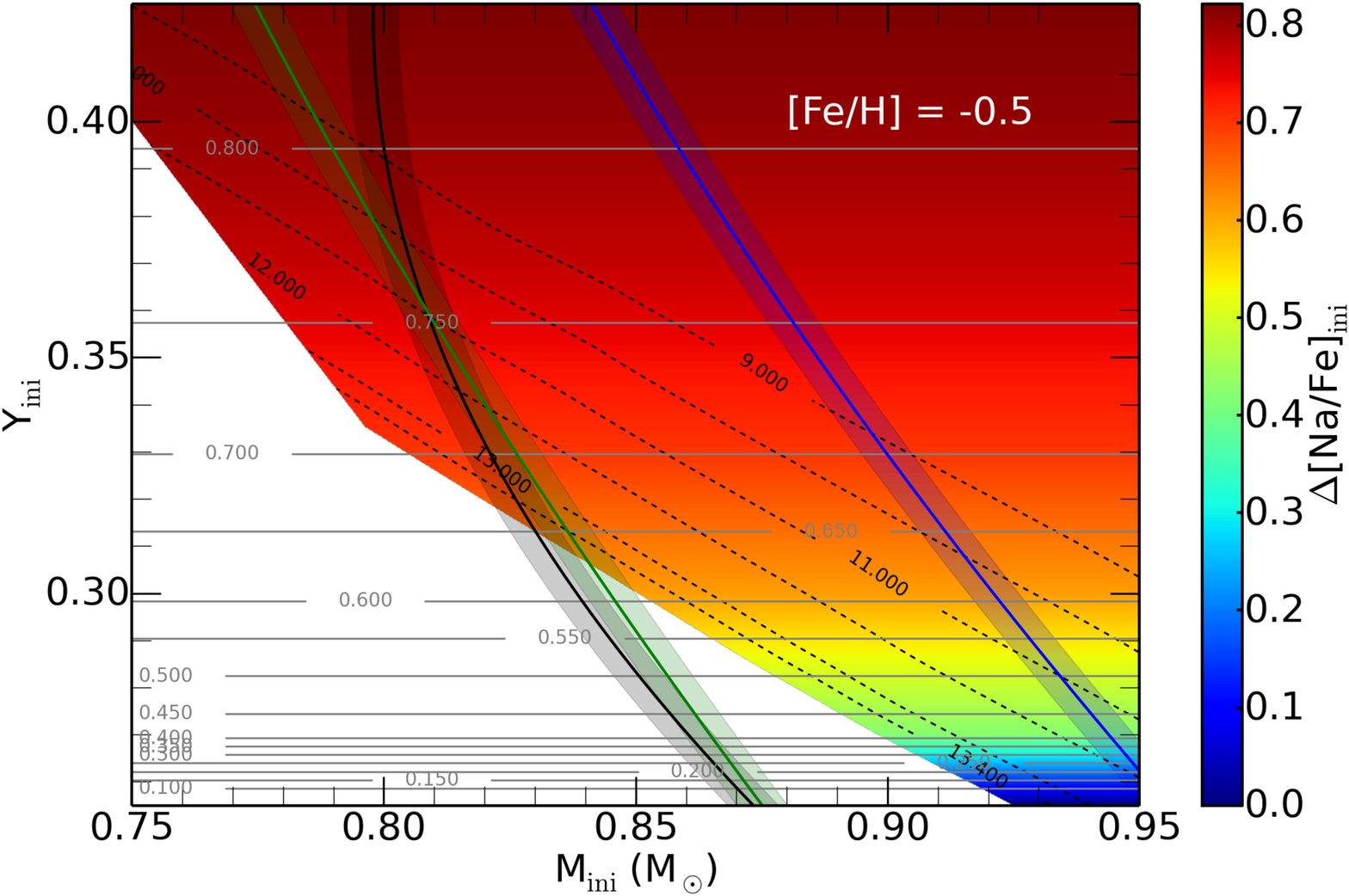} \\ 
\includegraphics[width=0.87\textwidth]{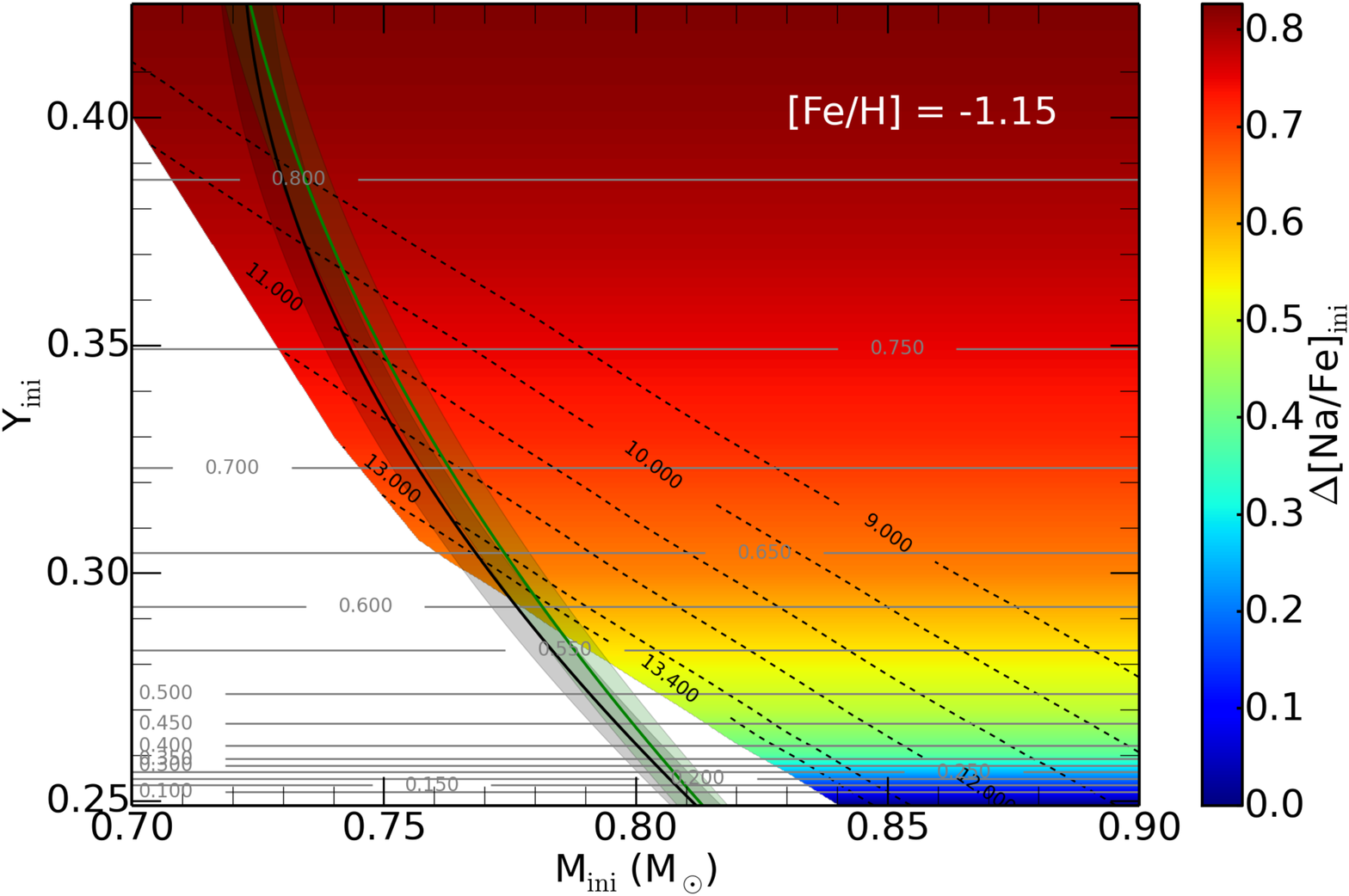} 
\end{center}
\caption{$\Delta$[Na/Fe]$_\mathrm{ini}$ (colour-coded) as a function of initial stellar mass and initial helium mass fraction for the highest [Fe/H] values considered in this study. Grey horizontal lines denote constant values of Na enrichment $\Delta$[Na/Fe]$_\mathrm{ini}$. 
The black, green, and blue full lines separate the stars that climb the AGB (to the right of the lines) from those that do not (left), with the shaded areas corresponding to error bars of 0.005~M$_{\odot}$ related to the mass step of the model grid. 
The black limit corresponds to the most stringent definition of the AGB (a star undergoes at least two thermal pulses), while the green and blue curves are for the C13 definition of the AGB (i.e., early-AGB only is included). The black and green cases on the one hand and the blue case on the other hand are obtained assuming a value of $\eta$ of 0.5 and 0.65, respectively, in the Reimers mass loss prescription. 
The black dashed lines are isochrones representing the ages at  the end of the helium-burning (in Gyr). The white area corresponds to the domain where stellar models have an age higher than 13.8 Gyr at the end of the helium-burning phase or do not enter that phase}
\label{figure:deltaNacolorvariousFeH}
\end{figure*}

\begin{figure*}[ht]
\begin{center}
\includegraphics[width=0.87\textwidth]{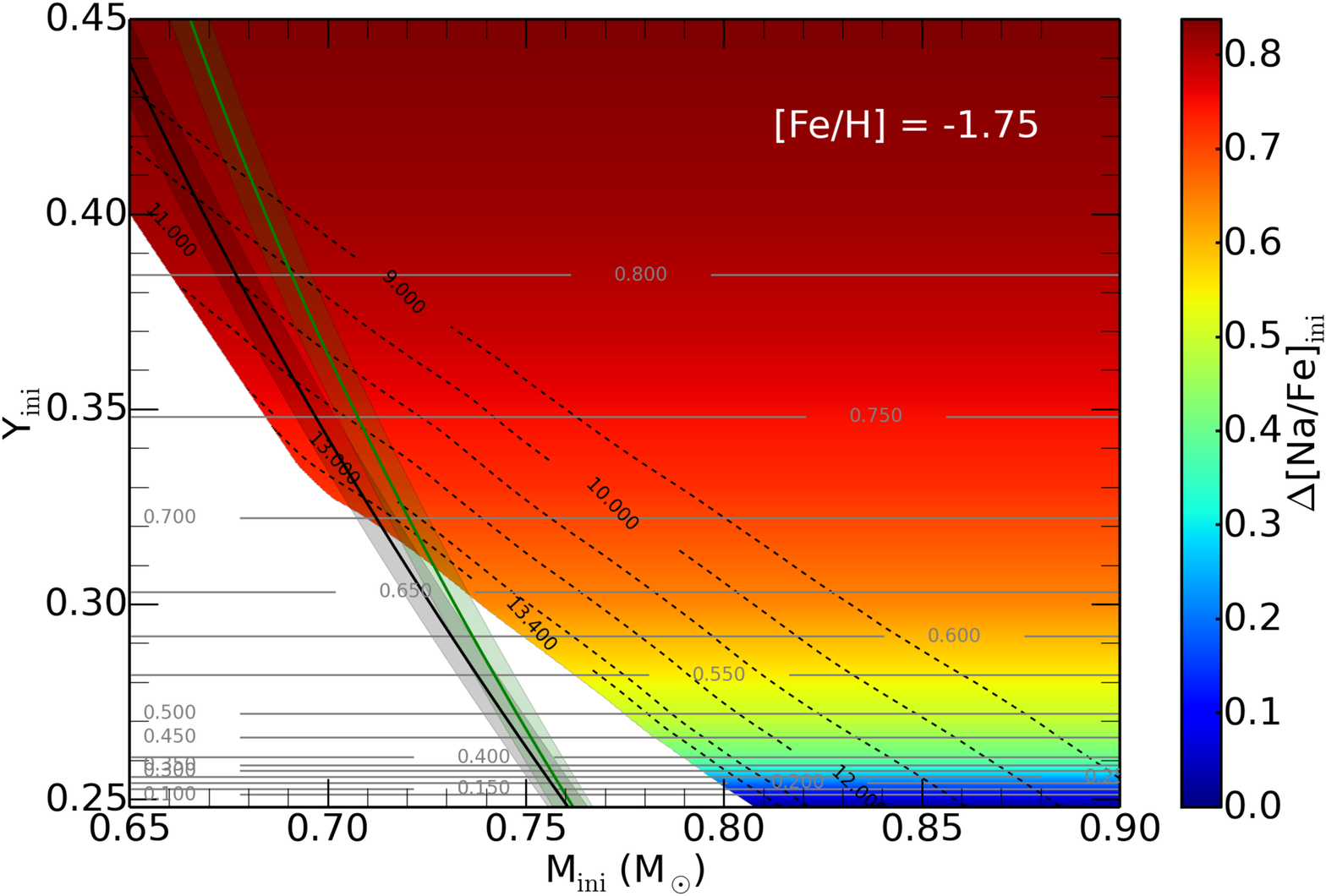} \\
\includegraphics[width=0.87\textwidth]{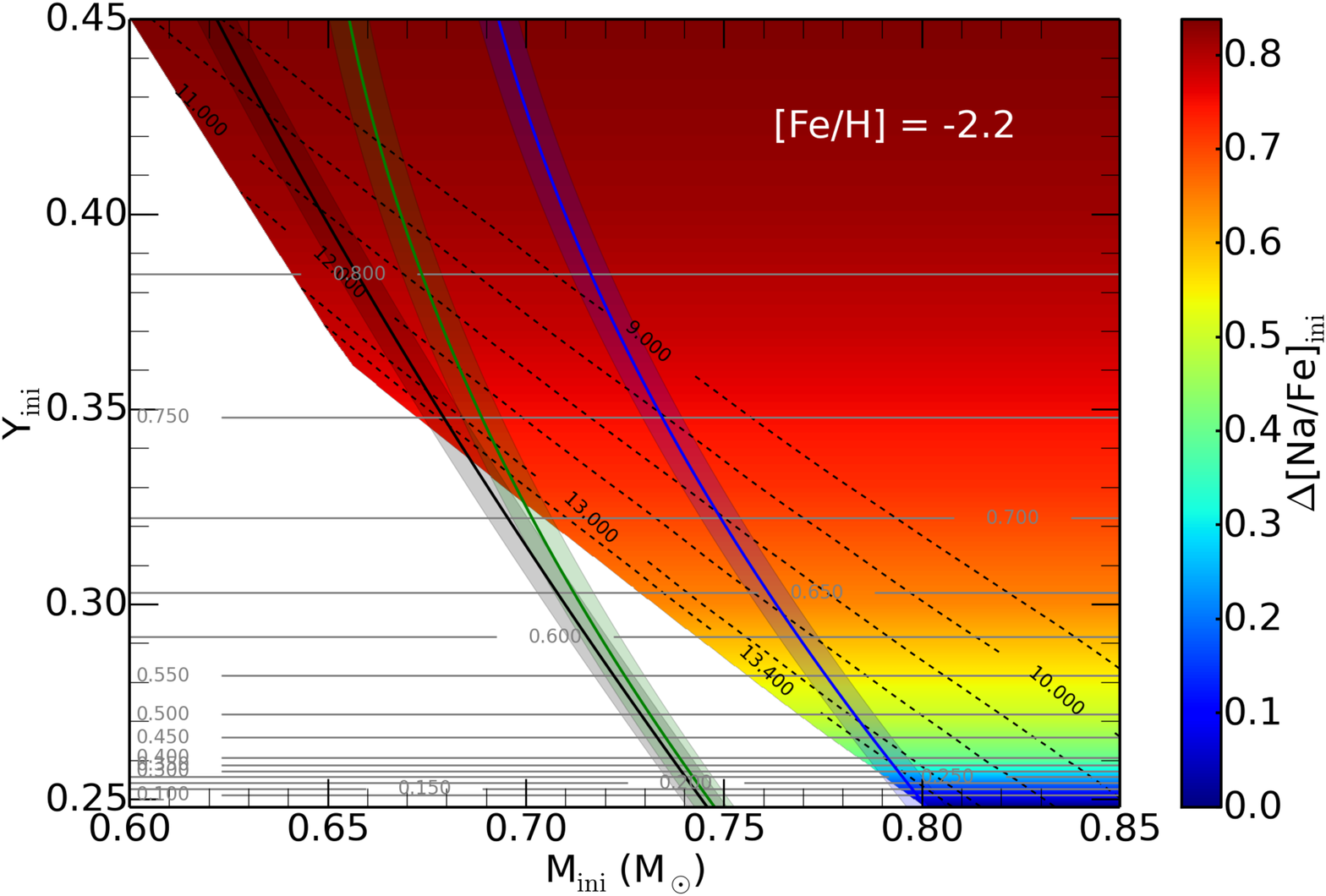}
\end{center}
\caption{Same as Fig.~\ref{figure:deltaNacolorvariousFeH} for the lowest [Fe/H] values considered in this paper}
\label{figure:deltaNacolorvariousFeH2}
\end{figure*}

This leads to the AGB / no-AGB limit curves shown in Figs.~\ref{figure:deltaNacolorvariousFeH} and \ref{figure:deltaNacolorvariousFeH2} for the different metallicities;
the dashed areas correspond to the typical error bar of 0.005~M$_{\odot}$ that is due to the mass step of the grid of models. The black curve is for the case where we consider as AGB stars those that undergo at least two thermal pulses, and the green curve is for the C13 case where stars make an excursion on the early-AGB without necessarily undergoing TP later on (shown only for [Fe/H] = -2.2 and -0.5). 
Both the black and green cases are for models computed with a value of the parameter $\eta$ equal to 0.5 in the Reimers prescription for mass loss on the RGB.
Finally, the blue curve and blue shaded area in the [Fe/H] = -2.2 and -0.5 plots correspond to the AGB limit (C13 criterion) when $\eta$ is taken equal to 0.65.
Clearly, variations of the mass loss parameter lead to an important shift of the AGB / no-AGB limit, as expected; however the definition of the AGB (TPs or not) has only a very modest impact.

The intersection of the AGB / no-AGB curves with the isochrones plotted in Figs.~\ref{figure:deltaNacolorvariousFeH} and \ref{figure:deltaNacolorvariousFeH2} provides the maximum (Y$_{\rm ini}$; M$_{\rm ini}$) one might expect for an AGB star at a given age for each value of [Fe/H]. 
For the three lowest metallicities considered ([Fe/H] = -1.15, -1.75, and -2.2), the cut-off is found at higher Y$\mathrm{ini}$ value 
for both younger ages at a given [Fe/H] and lower metallicities at a given age. 
The [Fe/H] = -0.5 curves does not follow the general trend; indeed in this case and for the mass range considered, the cut-off depends more on the 
 initial stellar mass than on the initial helium content.

\clearpage
\newpage

\clearpage
\newpage

\subsection{Theoretical sodium spread on the AGB as a function of metallicity and age, and limitations due to mass loss uncertainties}
\label{subsec:NaspreadvsZage}

The intersection of the AGB / no-AGB curves with the isochrones and the iso-$\Delta$[Na/Fe]$_\mathrm{ini}$ curves
in Fig.~\ref{figure:deltaNacolorvariousFeH} and \ref{figure:deltaNacolorvariousFeH2} allow us to also derive the maximum Na content expected on the AGB as a function of GC age, for the four [Fe/H] values considered. In these figures the colour code corresponds to the sodium enrichment of the 2P models with respect to the sodium abundance of 1P stars of similar metallicity.  

From this, we obtain the maximum helium and sodium enrichment of the 2P models that are expected to lie on the AGB (namely, $\Delta$ \Yini (AGB) and  $\Delta$[Na/Fe]$_\mathrm{ini}$(AGB)) for GCs of different ages and metallicities. This information is plotted in Fig.~\ref{fig:YNa_AGBvsageandFesurH}. 
The combination of effects on stellar evolution due to helium and metallicity described above together with the correlation between helium and sodium initial enrichment we adopt following the FRMS scenario (\S~\ref{subsec:secondpopmodels}) lead to the following results: 
\begin{itemize}
\item At a given metallicity, younger clusters are expected to host AGB stars exhibiting a larger helium and sodium spread than older clusters. 
\item At a given age, higher helium and sodium dispersion along the AGB is predicted in the more metal-poor GCs than in the metal-rich ones.
\item Both correlations hold for the two values of $\eta$ used in the Reimers prescription for mass loss.  The $\Delta$[Na/Fe]$_\mathrm{ini}$(AGB)-age slope is actually steeper for higher mass loss rates, and it increases with increasing metallicity.
\end{itemize}

More quantitatively, for $\eta$=0.5 and according to Fig.~\ref{fig:YNa_AGBvsageandFesurH}, the maximum and minimum sodium spreads expected among AGB stars ($\Delta$[Na/Fe]$_\mathrm{ini}$(AGB)), are 0.82~dex at 9 Gyr for [Fe/H]~=~-2.2, and 0.62 at 13.4 Gyr for [Fe/H]~=~-1.15, respectively. Since the maximum initial sodium enrichment predicted by the FRMS scenario is $\Delta$[Na/Fe]$_\mathrm{ini}$ = 0.86 (this is also the maximum theoretical dispersion on the RGB), we would therefore expect a difference of sodium dispersion between the RGB and AGB between 0.04 dex and 0.25 dex. 
Since the typical error bar for Na abundance determination on the AGB is of the order of 0.14 dex according to \cite{Campbell13}, the AGB-RGB difference could be detectable only for the old Galactic GCs of intermediate metallicity. Otherwise,  the observed [Na/Fe] dispersion on the AGB would be similar to that found on the RGB, as in NGC 104 \citep{Johnsonetal15_47Tuc} and NGC~2808 (Wang et al., in prep.), as discussed in \S~\ref{section:conclusion}.

As shown in Figs.~\ref{figure:deltaNacolorvariousFeH},  \ref{figure:deltaNacolorvariousFeH2},  and \ref{fig:YNa_AGBvsageandFesurH}, the cut-off limit in initial helium and sodium abundance and on the stellar mass at which a star climbs the AGB is very sensitive to the mass loss rate adopted along the RGB. This was discussed at length in C13 for the case of NGC~6752\footnote{\citet{Campbell13} proposed fitting the sodium distribution along the AGB in NGC~6752 by assuming an increased mass loss rate by a factor of 20 on the HB for all 2P stars. This proposal, which would require mass loss rates of the order 10$^{-9}$~M$_{\odot}$ yr$^{-1}$ (i.e., significantly higher than current observational and empirical constraints) was carefully and critically discussed by \citet{Cassisietal2014}.}.  
With the present grid of models we are able to quantify the impact of mass loss rate variations when [Fe/H] changes, and find that it actually carries the most important uncertainty when it predicts Na spread on the AGB at a given age and metallicity.
Indeed, a change from 0.5 to 0.65 for $\eta$ drastically decreases the Na cut-off values, especially when age increases. 
For the youngest GCs, this change of prescription does not have a dramatic impact. At 9~Gyr, the sodium cut-off value decreases by 0.03 dex for [Fe/H] = -2.2 and by 0.1 dex for [Fe/H] = -0.5 when $\eta$ increases from 0.5 to 0.65. 
However, at 13.4 Gyr the cut-off value decreases by 0.36 dex for the lowest considered metallicity for the same variation in $\eta$. 
For the highest metallicity this cut-off value drops even more quickly with increasing age when $\eta$=0.65, making the highest mass loss parameter unrealistic for metal-rich and old GCs, given that above 12 Gyr no more AGB stars are expected. 
This agrees well with the conclusions of \citet{McDonald15}, who derive a median value of 0.477$\pm$0.070 for $\eta$ from a sample of 56 well-studied Galactic GCs, and find only a weak $\eta$ gradient with metallicity although random spread exists among clusters at any metallicity.

\subsection{Uncertainties on the dependency of the initial He-Na correlation with metallicity}

In the FRMS scenario, 2P stars are expected to form in the immediate vicinity of 1P massive polluters. Therefore, their composition must bear the chemical signatures of their parent ``donor'' and its local ISM environment. 
In the present study we assume that the He-Na correlation in the initial abundance of 2P stars is identical at all metallicities.
More precisely, we use the abundance prescriptions derived from the H-burning ashes of the 60 and 120~M$_{\odot}$ FRMS models at [Fe/H]=-1.75 of \citet{Decressin07a} taking into account dilution with pristine gas so as to reproduce the Li-Na anticorrelation observed in NGC~6752. However, the internal temperature of massive stars is modified when metallicity changes, which affects the temperature of the CNO cycle and of the NeNa and MgAl chains. As a consequence, one might expect slightly different correlations between He and Na within the nucleosynthetic yields of 1P polluters of different metallicities. Also, more massive polluters might play a role in the self-enrichment of more metal-poor GCs, which might also slightly affect the extent of the He-Na correlation for 2P stars. These effects will have to be quantified in the future when FRMS models are available for the full metallicity range of Galactic GCs.

Here it is important to emphasize that despite these quantitative uncertainties, the fact that sodium and helium abundances are correlated for 2P stars is a strong prediction of the FRMS scenario. Instead, in the AGB scenario the helium enrichment of 2P stars is limited (maximum \Yini of $\sim$ 0.36, \citealt{Doherty14} as a result of second dredge-up in intermediate-mass stars), and it is not directly correlated with the sodium enrichment. In this case, therefore, the Na abundance distribution is expected to be the same on the RGB and AGB whatever the age and the metallicity of the cluster. This is at odds with the observations in NGC~6752 (see below).

\section{Comparison with observations and conclusions}
\label{section:conclusion}

\begin{figure*}
\begin{center}
\includegraphics[width=.495\textwidth]{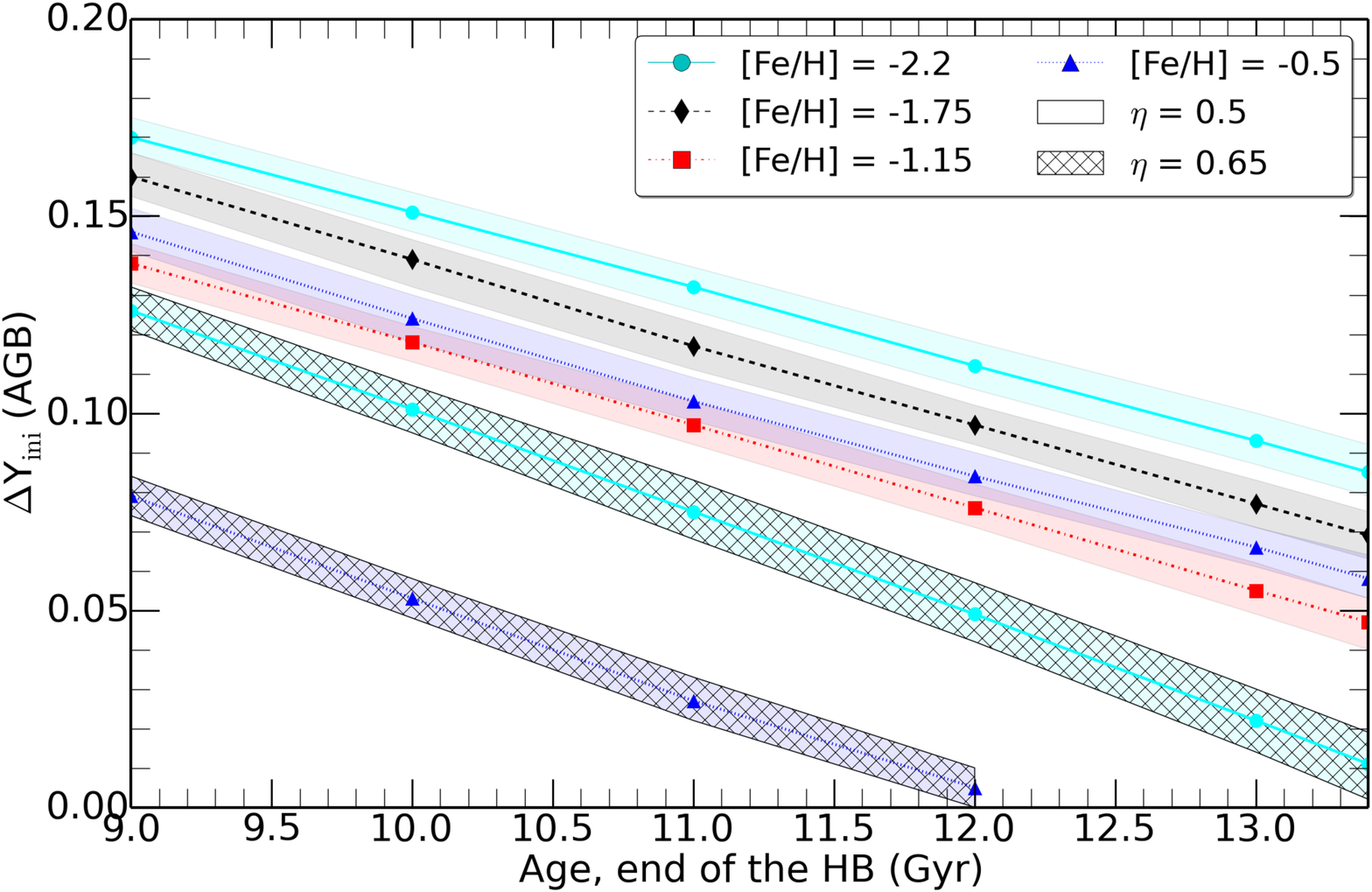} 
\includegraphics[width=.495\textwidth]{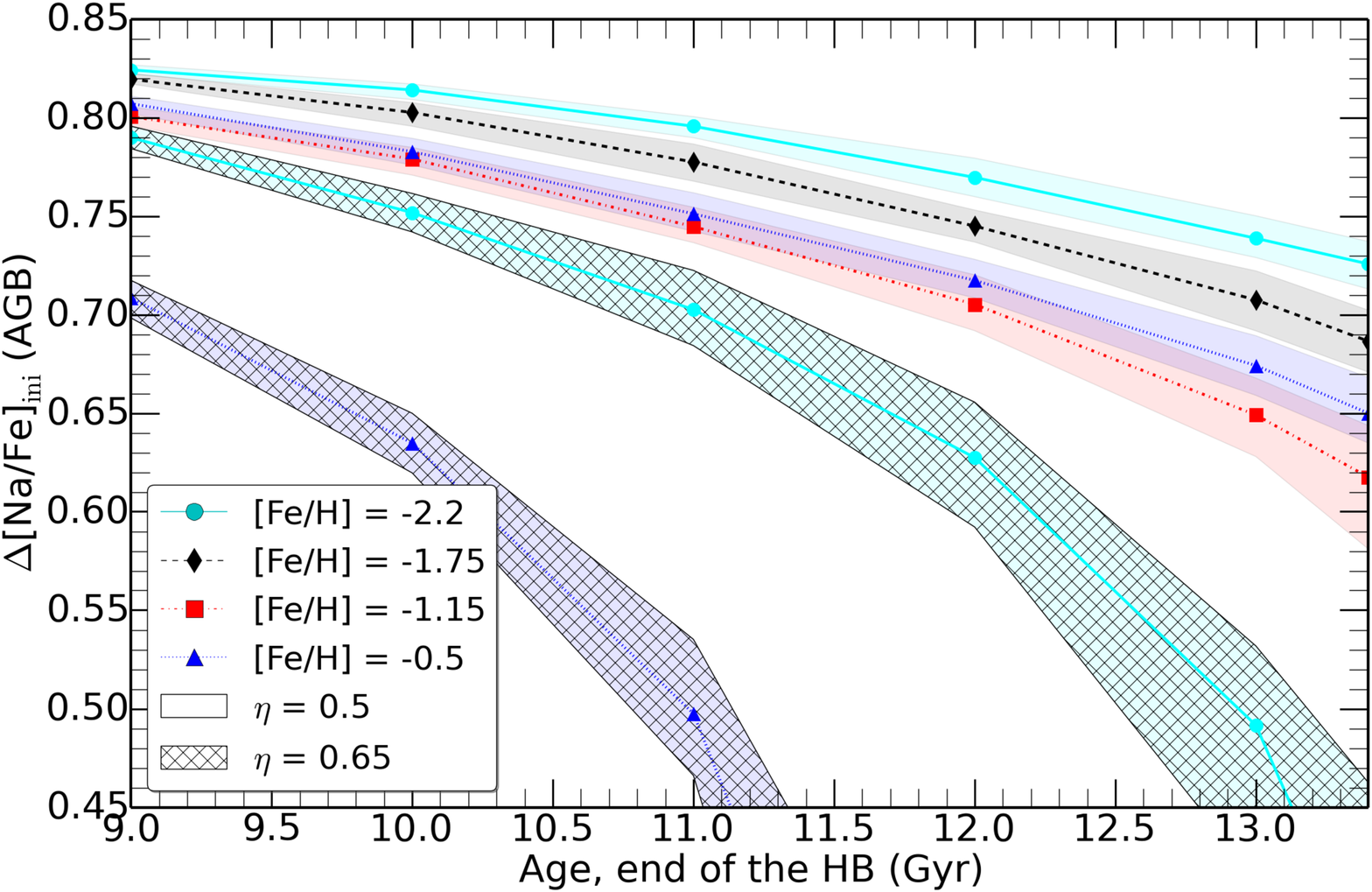}
\end{center}
\caption{Maximum spreads in helium and sodium (namely $\Delta$Y$_\mathrm{ini}$(AGB) and $\Delta$[Na/Fe]$_\mathrm{ini}$(AGB)) predicted on the AGB for the four metallicities as a function of the age at the end of the helium-burning phase. The shaded areas are obtained for a value of $\eta$ = 0.65 (shown only for the two extreme [Fe/H] values), while the other coloured areas are for $\eta$ = 0.5 for the four [Fe/H] values considered (see inset)}
\label{fig:YNa_AGBvsageandFesurH}
\end{figure*}

The trends with age and metallicity described above are based on well-understood stellar physics and are robust qualitatively. 
However, a number of limitations have to be kept in mind when it comes to quantitatively comparing model predictions with observations.  As shown in detail, quantitative results for the Na abundance distribution on the AGB strongly depend on the adopted mass loss rate. Uncertainties on the initial He-Na correlation may also vary from cluster to cluster and with metallicity. 
Another difficulty to performing quantitative comparison between models and observations comes from the fact that the ages quoted in the literature for GCs are derived with different stellar models to those computed for the present study. Therefore, in Fig.~\ref{fig:YNa_AGBvsageandFesurH} we refrain from plotting the sodium spread that has only been derived for a limited number of of GCs so far. However, we can note the following qualitative agreement with the trends we obtain. 

Model predictions by C13 obtained using FRMS prescriptions for the chemical composition of GC 2P stars were found to agree well (within theoretical and observational errors) with the lack of Na-rich stars discovered by \citet{Campbell13} in the specific case of NGC~6752. 
This cluster is relatively metal-poor ([Fe/H]=-1.54, \citealt{Campbell13}). It is also relatively old, as estimates of its age vary between $\sim12.5\pm$0.25~Gyr and 13.4$\pm$1.1~Gyr (according to \citealt{VandenBerg13} and \citealt{Gratton03} respectively). 
Therefore, its position in the age-$\Delta$[Na/Fe]$_\mathrm{ini}$(AGB) plane (Fig.~\ref{fig:YNa_AGBvsageandFesurH}) is located 
in the region where the theoretical value for maximum Na enrichment along the AGB drops sharply, and where we expect to see differences between RGB and AGB stars. 

In contrast, \citet{Johnsonetal15_47Tuc} find that the AGB population in 47~Tuc (NGC~104) presents a much larger [Na/Fe] dispersion, which is nearly identical to that found on the RGB. This GC is both more metal-rich ([Fe/H]=-0.73, \citealt{Johnsonetal15_47Tuc}) and younger (11.75$\pm$0.25, \citealt{VandenBerg13}). It would thus lie to the left of NGC~6752 in the age-$\Delta$[Na/Fe]$_\mathrm{ini}$(AGB) plane (Fig.~\ref{fig:YNa_AGBvsageandFesurH}). In this region, larger AGB  sodium dispersion is predicted than in the case of NGC~6752.

Finally, Wang et al. (in prep.) find a similar Na spread on the RGB and AGB of NGC~2808, a relatively young GC (11.00$\pm$0.38,  \citealt{VandenBerg13}) of intermediate metallicity ([Fe/H]=-1.17, \citealt{Wangetal2015IAUS316}). This cluster lies to the 
left of both NGC~6752 and 47~Tuc in the age-$\Delta$[Na/Fe]$_\mathrm{ini}$(AGB) plane (Fig.~\ref{fig:YNa_AGBvsageandFesurH}). At this location, RGB and AGB stars are expected to have tovery similar Na dispersion within observational errors.

We can therefore conclude that an initial He-Na correlation like the one assumed in the present study leads to a reasonable solution to different AGB sodium abundances found from cluster to cluster depending on their age and initial metal content.  However, cluster to cluster variations are expected, due in particular to mass loss and self-enrichment depending on intrinsic cluster properties (e.g., mass or compactness). This can be easily tested observationally by a detailed chemical analysis of AGB stars in GCs spanning a wide range in both age and metallicity. 

\begin{acknowledgements}
We acknowledge support from the Swiss National Science Foundation (FNS) for the project 200020-159543 ``Multiple stellar populations in massive star clusters – Formation, evolution, dynamics, impact on galactic evolution" (PI C.C.). We thank the International Space Science Institute (ISSI, Bern, CH) for welcoming the activities of ISSI Team 271 ``Massive star clusters across the Hubble Time'' (2013 - 2016, team leader C.C.). We thank the anonymous referee for the useful comments that helped us improve the presentation of our results.
\end{acknowledgements}

\bibliographystyle{aa}
\bibliography{Bibliography}

\end{document}